# Investigation of Sb diffusion in amorphous silicon

A. Csik[1,*], G.A. Langer[1], G. Erdélyi[1], D.L. Beke[1], Z. Erdelyi[1], K. Vad[2]

[1]*Department of Solid State Physics, University of Debrecen, P.O. Box 2, Debrecen H-4010, Hungary*

[2] *Institute of Nuclear Research, Hungarian Academy of Sciences (ATOMKI), P.O. Box 51, Debrecen H-4001, Hungary*

Amorphous silicon materials and its alloys become extensively used in some technical applications involving large area of the microelectronic and optoelectronic devices. However, the amorphous-crystalline transition, segregation and diffusion processes still have numerous unanswered questions. In this work we study the Sb diffusion into an amorphous Si film by means of Secondary Neutral Mass Spectrometry (SNMS). Amorphous Si/Si$_{1-x}$Sb$_x$/Si tri-layer samples with 5 at% antimony concentration were prepared by DC magnetron sputtering onto Si substrate at room temperature. Annealing of the samples were performed at different temperature in vacuum ($p<10^{-7}$ mbar) and 100 bar high purity (99.999%) Ar pressure. During annealing a rather slow mixing between the Sb-alloyed and the amorphous Si layers was observed. Supposing concentration independent diffusion, the evaluated diffusion coefficients are in the range of $\sim 10^{-21}$ m$^2$s$^{-1}$ at 823 K.

---

[*] Corresponding author. *Tel./Fax*: +36-52-316073
E-mail address: csik@atomki.hu (A. Csik)
*On leave from Institute of Nuclear Research, Hungarian Academy of Sciences (ATOMKI)*

**Introduction**

During the last years amorphous Si and a-Si based alloys have been widely studied and utilized in micro- and optoelectronic devices [1,2,3]. The new generation of these devices (thin film transistors, single electron transistor, etc.) require p- or n-type doped active channels with high level of concentration and concentration variation control. Antimony (Sb) is the most frequently used n-type dopant [4]. Thus the knowledge of diffusion mechanisms and data is highly desired for understanding the physical properties of this material. Although the diffusion of Sb in crystalline Si is widely studied, there are only few dates about diffusion in amorphous silicon. Study of diffusion in amorphous materials presents some difficulties. One of the main problems is to keep the thermal stability of the amorphous phase.

In our previous works [5,6] the diffusion and transformation processes in amorphous $Si_{1-x}Sb_x$ mono-, $Si/Si_{1-x}Sb_x/Si$ tri- and $Si/Si_{1-x}Sb_x$ multilayer systems with different composition (x=13-26 %) were investigated. After annealing the samples the most interesting result was that under hydrostatic pressure (100 bar-4,7 kbar Ar) in the Si/SiSb/Si trilayers (and only in this type of samples and only under hydrostatic pressures) the SiSb layer underwent a spinodal-like decomposition, resulting in three Sb-rich stripes parallel to the interfaces. The transformation at given temperature strongly depended on the initial Sb concentration of the SiSb layers and the applied hydrostatic pressure. No such stripes could be detected when the samples of the same geometry and composition were annealed in vacuum. The phenomenon was interpreted by means of an amorphous-nanocrystalline transformation (which enhanced the effective diffusivity) and by segregation of Sb at the interfaces. Antimony diffusion through an amorphous Si film and segregation kinetics on the top of the Si layer was also studied [7] in situ by means of Auger electron spectroscopy. The antimony diffusion coefficient proved to be 10-11 orders of magnitude higher than that of measured in crystalline Si.

In this article, we present additional experimental results on the diffusion of Sb in the amorphous Si/Si$_{1-x}$Sb$_x$/Si system. In contrast to our former experiments, the concentration of Sb in the samples was reduced to 5 % to avoid the fast crystallization and decomposition in the SiSb layer. Secondary Neutral Mass Spectrometry (SNMS) has been applied to measure the distribution of Sb in as-deposited and heat treated films. In order to detect the possible crystallization and structural transformation processes of the samples Transmission Electron Microscope (TEM) measurements were carried out. The diffusion coefficient of Sb was calculated from the time dependence of the experimentally determined antimony concentration distribution curves.

**Preparation and experimental method**

Amorphous samples with the structure Si(20nm)/Si$_{95at\%}$Sb$_{5at\%}$(20nm)/Si(20nm) were prepared by alternate sputtering of pure elements Si(99.999at%) and Sb(99.99at%) in a dc magnetron system with $5 \cdot 10^{-1}$ Pa Ar pressure during deposition. The base pressure of the system was $10^{-5}$ Pa. The composition of the SiSb layer was adjusted by covering of the Sb target with Si plates. The Sb content of the SiSb film was determined by energy-dispersive X-ray spectrometer (Oxford Link-Isis EDS). The samples were sputtered at room temperature on Si(100) substrates. Annealing treatments were performed at temperatures of 723 K, 773 K, 823 K, 873 K and 1023 K with durations ranging from 0.5 to 10 hours in vacuum (<$10^{-5}$ Pa) and in 100 bar high purity Ar pressure.

Microstructural characterization of the samples was carried out using a JEOL type Transmission Electron Microscope (TEM) and an X-ray diffraction instrument installed with a Siemens Cu-anode X-ray tube. The concentration profiles were measured by Secondary Neutral Mass Spectrometer (SPECS INA-X). Antimony presence on the surface of samples was investigated by X-Ray Photoelectron Spectroscopy (XPS, ESA 31 – ATOMKI).

**Results and discussion**

The cross-sectional TEM analysis proves that the as-deposited layer structures are well defined and perfectly planar (Fig. 1.). The antimony concentration of the alloyed layer was reduced to about 5%, in order to avoid the crystallization and decomposition in the samples. The SNMS measurements showed that no concentration changes take place during the annealing of the series of samples in vacuum at temperatures 723 K, 773 K and 823 K for different times (Fig. 2.). This result is in agreement with our previous work [5], where the samples with higher Sb concentration (18 and 24 at%) annealed in vacuum also showed no detectable changes. From these experimental results it can be concluded that no measurable changes can be detected in Sb concentration profile at temperature range 723 – 823 K after 10 hours annealing.

On the other hand, applying 100 bar Ar pressure during the annealing, the Sb concentration profile of the heated samples shows broadening (Fig. 3.), which means that Sb diffused out from SiSb layer into amorphous Si. However, in contrast to heated samples in 100 bar Ar with higher Sb concentration (13-26 at%) [6], where the SiSb layer underwent a spinodal-like decomposition resulting three Sb-rich stripes parallel to the original interfaces, morphological changes were not detected. Electron diffraction measurements also showed that the samples heat treated at 823 K remained in amorphous state.

The surface of the heated samples was investigated by XPS. In these measurements the presence of Sb at the surface and next to the surface was not detectable.

The concentration and time independent diffusion coefficient of the Sb was extracted from the broadening of the concentration profile using the following procedure. The antimony concentration profile of the as-grown samples measured by SNMS was approximated by the function:

$$c(x, t = 0) = A \cdot [erf((a - x)/c) + erf((x - b)/c)]$$

where *A, a, b, c* are constants and estimated from the best fit. The Fick-equation was solved numerically using the partial differential equation solver of the MatLab (Fig. 4.). The diffusion fluxes of the Sb atoms were supposed to be zero at the top surface and at the substrate/a-Si surface. The calculated diffusion coefficient of Sb at 823 K was found to be $\sim 1 \cdot 10^{-21}$ m$^2$/s.

To get information for the temperature dependence of diffusion coefficient we carried out the annealing of the same samples at higher temperatures range (873 – 1023 K). Unfortunately, as it can be seen from TEM pictures that the samples annealed at 873 K in vacuum and under 100 bar Ar pressure for 5 hours become crystalline (Fig.5., 6.). It can be clearly seen that small crystals appears at 873 K and sample annealed 150 K higher can shows fully crystallized sate. These results are in accordance with the findings of Carlsson et al. [8]. They showed that the sample with the same Sb concentration after 30 minutes annealing at 1023 K still remains in amorphous state. We suppose that at the beginning of the annealing the Sb diffuses across amorphous Si layer and later due to the metal-induced crystallization (MIC) effect [8,9] starts to crystallize.

**Conclusion**

We have used TEM and SNMS to investigate the antimony diffusion in amorphous Si. It was showed that sample with 5% Sb concentration annealed lower then 823 K remains in amorphous state. From the time evaluation of SNMS concentration profile $D_{Sb} \approx 1 \cdot 10^{-21}$ m$^2$/s diffusion coefficient was calculated for Sb diffusion in amorphous Si at 823 K. As it was described in our present and previous works [5,6] in annealed samples the diffusion and decomposition processes significantly enhanced by the applying small (100 bar) hydrostatic pressure. The effect of hydrostatic pressure is not clear and need further investigation.

**Acknowledgements**

This work was supported by OTKA Grant No. D-048594, T-043464, F-043372, K-61253. Z. Erdélyi acknowledges support from Bolyai János Foundation.

**Figure captions**

Figure 1. Cross-sectional TEM image of Si [20nm]/ Si$_{95\%}$Sb$_{5\%}$ [20 nm]/Si [20nm] as-depsited trilayer

Figure 2. Time evaluation of Sb concentration profile during annealing in vacuum (a.) and in 100 bar Ar (b.)

Figure 3. Fitted concentration profiles of Sb for the initial and heat treated states

Figure 4. TEM image of the sample after 5 hours heat treatment at 823 K

Figure 5. Fully crystallized Si/Si$_{95\%}$Sb$_{5\%}$/Si sample annealed at 1023 K for 1 hour

Figure 1. A. Csik, G.A. Langer, G. Erdélyi, D.L. Beke, Z. Erdelyi, K. Vad

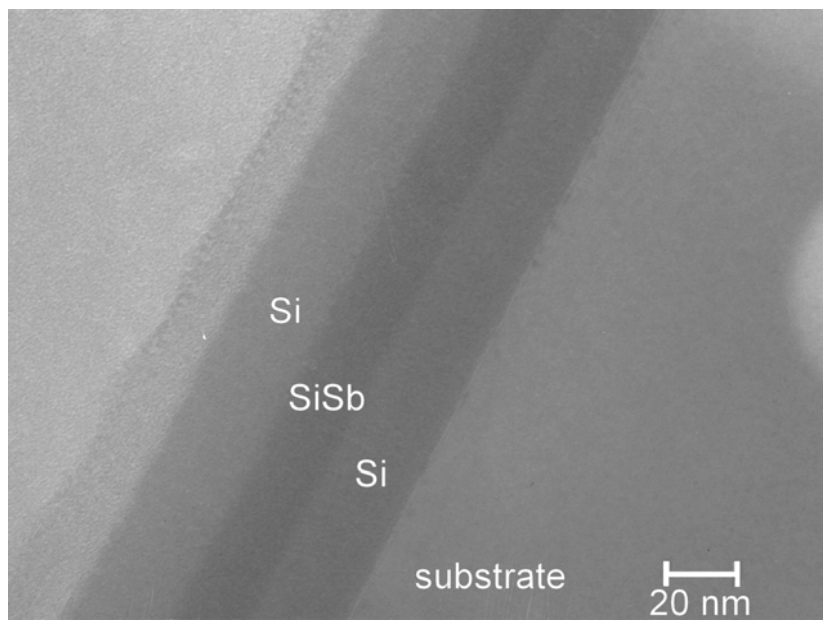

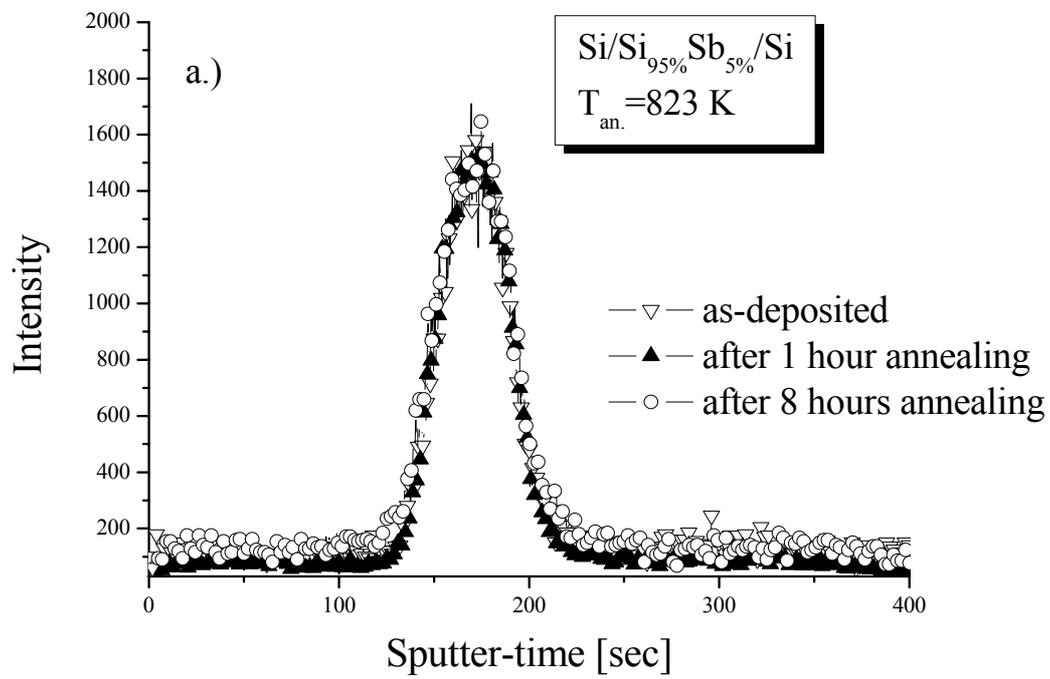

a.)

Si/Si$_{95\%}$Sb$_{5\%}$/Si
T$_{an.}$=823 K

—▽— as-deposited
—▲— after 1 hour annealing
—○— after 8 hours annealing

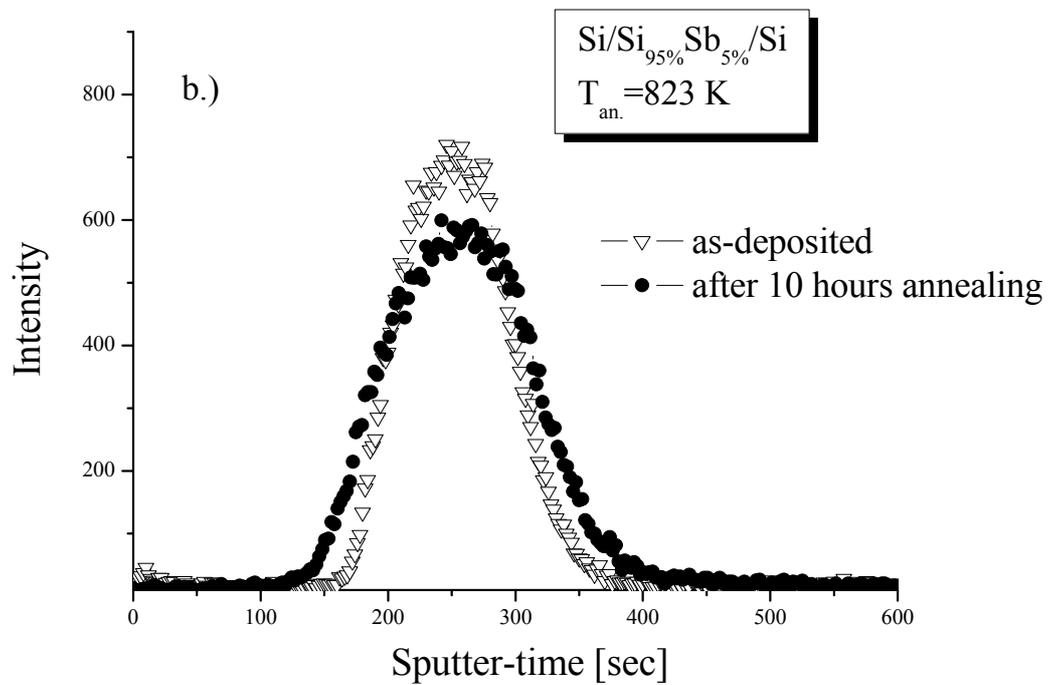

b.)

Si/Si$_{95\%}$Sb$_{5\%}$/Si
T$_{an.}$=823 K

—▽— as-deposited
—●— after 10 hours annealing

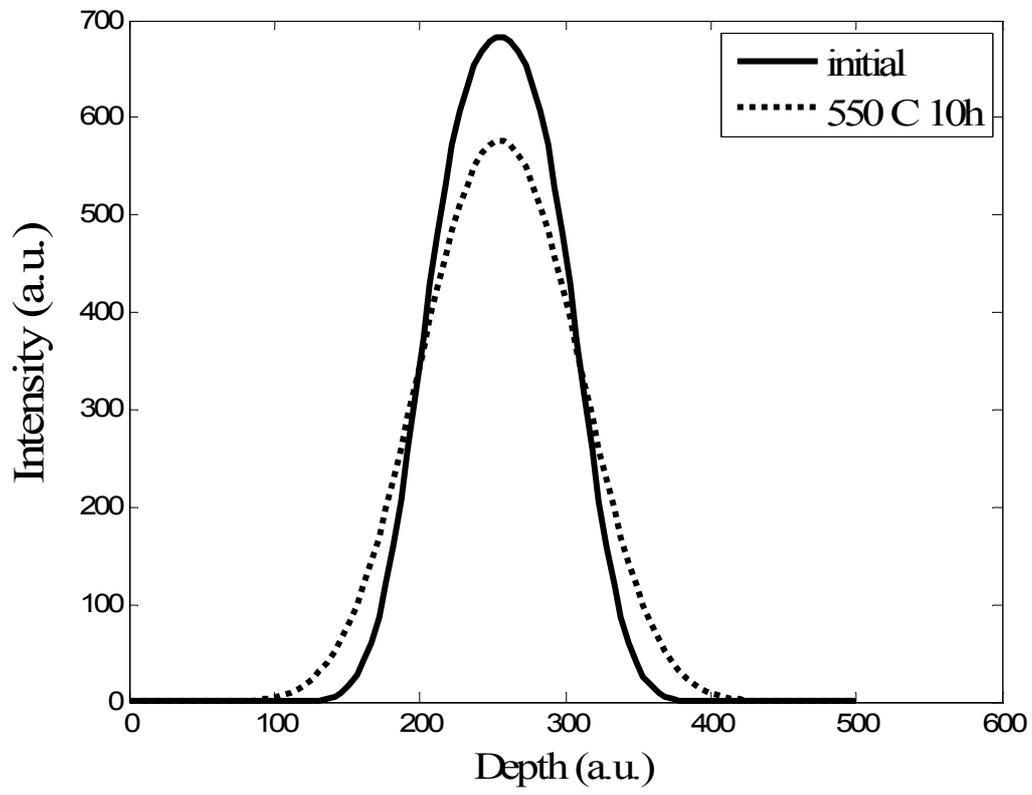

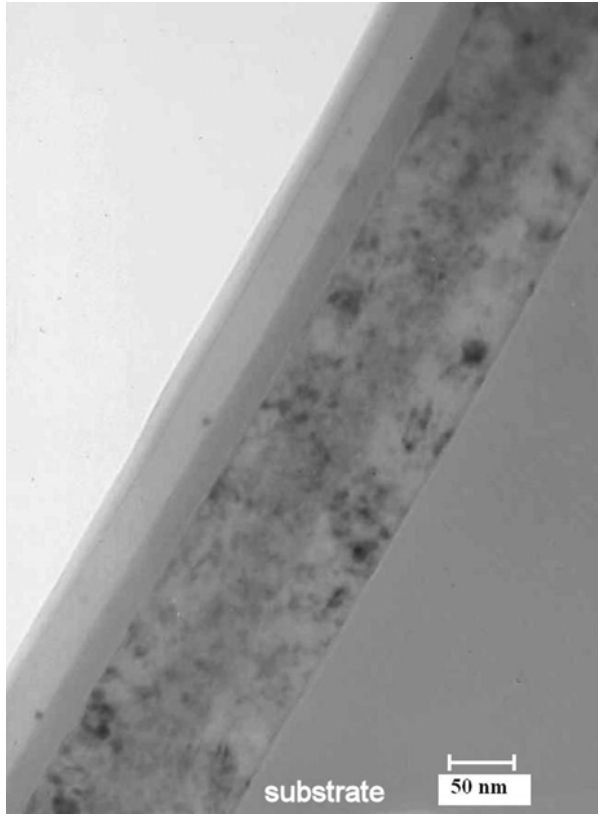

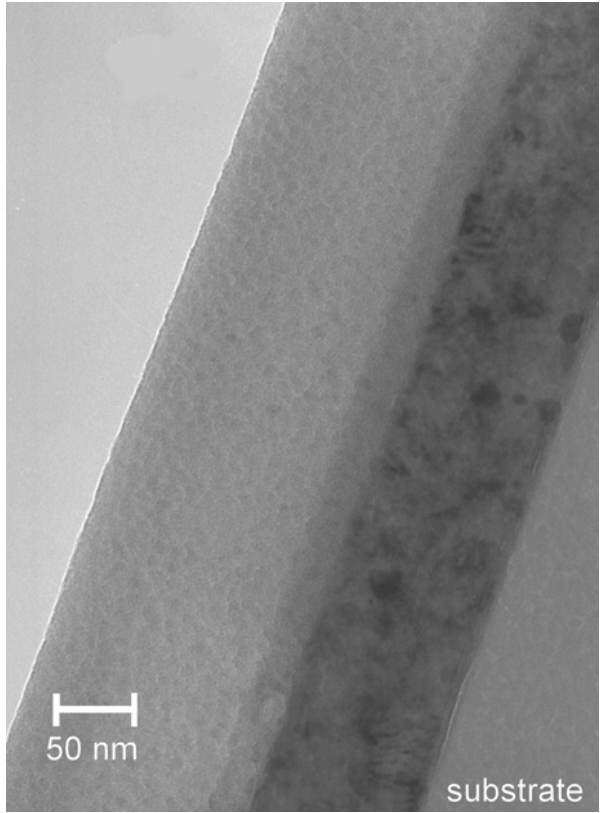